\documentstyle[prd,aps,epsf,floats,amsfonts,amssymb,amsmath]{revtex}

\newcommand{\be}{\begin{equation}}\newcommand{\ee}{\end{equation}}
\newcommand{\bea}{\begin{eqnarray}}\newcommand{\eea}{\end{eqnarray}}
\newcommand{\brr}{\begin{array}}\newcommand{\err}{\end{array}}
\newcommand{\bit}{\begin{itemize}}\newcommand{\eit}{\end{itemize}}
\newcommand{\ben}{\begin{enumerate}}\newcommand{\een}{\end{enumerate}}
\newcommand{\bib}{\bibitem}

\def\lab{\label}
\def\lan{\langle}
\def\lf{\left}

\def\non{\nonumber}
\def\ran{\rangle}

\def\ri{\right}

\def\al{\alpha}\def\bt{\beta}
\def\Ga{\Gamma}
\def\te{\theta}

\def\si{\sigma}

\def\1{{_{1}}}\def\2{{_{2}}}
\def\nof{:\;\!\!\;\!\!:}
\def\wQ{Q}
\def\wwQ{Q}
\begin{document}


\title{Lepton charge and neutrino mixing in pion decay processes}

\author{Massimo Blasone${}^{\flat }$, Antonio
Capolupo${}^{\flat}$,  Francesco Terranova${}^{\sharp}$ and
Giuseppe Vitiello${}^{\flat}$}

\vspace{5mm}

\address{${}^{\flat}$Dipartimento di Fisica and INFN,  Universit\`a di Salerno,
I-84100 Salerno, Italy \\ ${}^{\sharp}$Laboratori Nazionali
di Frascati dell'INFN, I-00044, Frascati, Italy}


\maketitle

\vspace{8mm}

\begin{abstract}
We consider neutrino mixing and oscillations in quantum field
theory and compute the neutrino lepton charge in decay processes
where neutrinos are generated. We also discuss the proper definition of flavor charge
and states and clarify the issues of the possibility of different mass parameters in field mixing.
\end{abstract}

\vspace{8mm}


\section{Introduction}

In the context of Quantum Field Theory (QFT) a rich
non--perturbative vacuum structure associated with the mixing of
fermion and boson fields has been revealed
\cite{BV95,BHV98,currents,Blasone:1999jb,Fujii:1999xa,binger,BCRV01,JM01,JM011,yBCV02,spaceosc,Blaspalm,Capolupo:2004av,Capolupo:2004pt,nulorentz}
and the exact formulas for
fermion and boson field oscillations \cite{BHV98,BCRV01,yBCV02,spaceosc,Blaspalm} are
now established. In particular, in a full QFT treatment particle
mixing exhibits new features with respect to the usual formulae in
Quantum Mechanics
QM \cite{chengli}. The phenomenological analysis for meson
mixing has shown that, while for most of mixed systems the
non-perturbative structure of the vacuum produces negligibly small
effects, for strongly mixed systems such as $\omega - \phi$ or
$\eta - \eta ~'$ non-perturbative corrections can be as large as
$5 -20\%$ \cite{Capolupo:2004pt}.
Moreover the non-perturbative field theory
effects may contribute in a crucial way in other physical
contexts. For example, as shown in  \cite{Blasone:2004yh},
the neutrino mixing
may contribute to the value of the cosmological constant exactly
because of the non-perturbative effects.

There are, however,
several aspects which still need to be fully developed.
For example, how to deal, in the presence of mixing,
with those decay processes where neutrinos are generated.

Since the time of the introduction of the Pontecorvo
mixing transformations \cite{chengli}, it is
well known that the mixed (flavor)
neutrinos  are not mass eigenstates. This
implies that flavor neutrinos are not representations of the
Poincar\'e group: one cannot think of them as asymptotic fields in
the frame of the Lehmann-Symanzik-Zimmerman (LSZ) formalism
\cite{Itz}. The QFT analysis of the mixing phenomenon has
indeed clarified \cite{BV95} that flavor neutrino field
operators do not have the mathematical characterization necessary
in order to be defined as asymptotic field operators acting on the
massive neutrino vacuum. The origin of this is related to the fact
that the vacuum for the massive neutrinos turns out to be
unitarily inequivalent to the vacuum for the mixed neutrino
fields.

 Previous works were mainly
focused on the determination of the oscillation probability
through the analysis of the expectation value of the flavor charge
operator
\begin{equation}
\langle \nu_\rho(t) | Q_{\nu_{\sigma}} | \nu_\rho(t) \rangle
\label{equ:Q_nu}
\end{equation}
$Q$ being a function of the flavor annihilation/creation operators
\cite{Lurie} and $\rho,\sigma=e,\mu,\tau$  flavor indices. In fact,
flavor states are produced in weak interaction processes and we
are left to the question whether Eq.(\ref{equ:Q_nu}) can be
consistently extended to include the neutrino production vertex
and what is the explicit form of
\begin{equation}
\langle \Pi(t) | Q_{\nu_{\sigma}} | \Pi(t) \rangle
\label{equ:Q_nu_vertex}
\end{equation}
where $| \Pi(t) \rangle$ represents the evolution at time $t$ of the parent
state for the neutrino.

The computation of the matrix element (\ref{equ:Q_nu_vertex}) is not trivial matter
since it simultaneously involves LSZ states and flavor states, which are not
LSZ.
This paper is devoted to the study of such matrix element
and of several related topics.
In Section II we review the formalism of neutrino mixing in QFT and discuss the proper definition
of flavor charges and states; in Section III we clarify the issues of the possibility of different mass parameters
in field mixing, which has recently attracted some attention \cite{Fujii:1999xa,BCRV01,JM011,yBCV02,Giunti:2003dg}.
In Section IV we perform a careful analysis of (\ref{equ:Q_nu_vertex}).
Conclusion are drawn in Section V.
For the reader convenience, we present in
Appendix A the proof of orthogonality of the flavor states at
different times. The Appendix B contains the explicit form of some equations derived
in the text.

\section{Flavor fields and states}

Let us start by introducing the general frame for our discussion,
which is also useful to set up our notations. For a detailed review see \cite{Capolupo:2004av}.
For simplicity we consider only two Dirac neutrino fields. The
Pontecorvo mixing transformations are \cite{chengli}
\begin{eqnarray} \nonumber
\nu_{e}(x)   &=&\nu_{1}(x)\,\cos\theta +
                        \nu_{2}(x)\,\sin\theta\\
\label{rot1} \nu_{\mu}(x) &=&-\nu_{1}(x)\,\sin\theta
                      + \nu_{2}(x)\,\cos\theta
\;,\end{eqnarray}
where $\theta$ is the mixing angle and $\nu_1$ and $\nu_2$ are
massive non-interacting, free fields, anticommuting with each
other at any space-time point. The fields $\nu_1$ and $\nu_2$ have
non-zero masses $m_{1} \neq m_{2}$ and are explicitly given by
\begin{equation}\label{2.2}
\nu_{i}(x) =
\frac{1}{\sqrt{V}} \sum_{{\bf k},r} \left[u^{r}_{{\bf k},i}(t)\;
\alpha^{r}_{{\bf k},i}\:+ v^{r}_{-{\bf k},i}(t)\;
\beta^{r\dag }_{-{\bf k},i} \right] e^{i {\bf k}\cdot{\bf x}}
, \; \; \; ~  i=1,2 \;.
\end{equation}
with $u^{r}_{{\bf k},i}(t)=e^{-i\omega_{k,i} t}u^{r}_{{\bf
k},i}(0)$, $v^{r}_{{\bf k},i}(t)=e^{i\omega_{k,i}
t}v^{r}_{{\bf k},i}(0)$ and  $\omega_{k,i}=\sqrt{{\bf
k}^2+m_i^2}$. Here and in the following we use $t\equiv x_0$, when
no misunderstanding arises. The vacuum for the $\alpha_i$ and
$\beta_i$ operators is denoted by $|0\rangle_{1,2}$: $\; \;
\alpha^{r}_{{\bf k},i}|0\rangle_{12}= \beta^{r }_{{\bf
k},i}|0\rangle_{12}=0$. The anticommutation relations are the
usual ones (see Ref.\cite{BV95}). The orthonormality and
completeness relations are:
\begin{equation}
u^{r\dag}_{{\bf k},i} u^{s}_{{\bf k},i} = v^{r\dag}_{{\bf k},i}
v^{s}_{{\bf k},i} = \delta_{rs},\qquad ~ u^{r\dag}_{{\bf k},i}
v^{s}_{-{\bf k},i} = v^{r\dag}_{-{\bf k},i} u^{s}_{{\bf k},i} = 0,
\qquad ~ \sum_{r}(u^{r}_{{\bf k},i} u^{r\dag}_{{\bf k},i} + v^{r}_{-{\bf
k},i} v^{r\dag}_{-{\bf k},i}) = 1.
\end{equation}

The fields $\nu_e$ and $\nu_\mu$ are completely determined through
Eq.(\ref{rot1}), which can be rewritten in the form (we use
$(\sigma,i)=(e,1) , (\mu,2)$):
\begin{equation} \label{exnue1}
\nu_{\sigma}(x) = G^{-1}_{\theta}(t) \nu_{i}(x) G_{\theta}(t) =
\frac{1}{\sqrt{V}} \sum_{{\bf k},r} \left[ u^{r}_{{\bf k},i}(t)\;
\alpha^{r}_{{\bf k},\nu_{\sigma}}(t) + v^{r}_{-{\bf k},i}(t)\;
\beta^{r\dag}_{-{\bf k},\nu_{\sigma}}(t) \right] e^{i {\bf k}\cdot{\bf
x}},
\end{equation}
with $G_{\theta}(t)$ the generator of the mixing transformations
Eq.(\ref{rot1})
\begin{equation}\label{BVgen}
G_{\theta}(t) = \exp\left[\theta \int d^{3}{\bf x}
\left(\nu_{1}^{\dag}(x)\nu_{2}(x) - \nu_{2}^{\dag}(x)
\nu_{1}(x)\right)\right].
\end{equation}

Eq.(\ref{exnue1}) provides an expansion of the flavor fields
$\nu_{e}$ and $\nu_{\mu}$ in the same basis of $\nu_{1}$ and
$\nu_{2}$. The flavor annihilation operators are then identified
with
\begin{equation}\label{BVoper}
\left(\begin{array}{c}
\alpha^{r}_{{\bf k},\nu_{\sigma}}(t)\\
\beta^{r\dag}_{{-\bf k},\nu_{\sigma}}(t)
\end{array}\right)
= G^{-1}_{\theta}(t) \left(\begin{array}{c}
\alpha^{r}_{{\bf k},i}\\
\beta^{r\dag}_{{-\bf k},i}
\end{array}\right)
G_{\theta}(t)
\end{equation}
The flavor vacuum is defined as $|0(t)\rangle_{{e,\mu}}\equiv
G^{-1}_{\theta}(t)|0\rangle_{1,2}\, $ and turns out to be orthogonal to the
vacuum for the mass eigenstates $|0\rangle_{1,2}$ in the infinite volume limit.
Note the time dependence of $|0(t)\rangle_{{e,\mu}}$: it turns out that flavor
vacua taken at different times are orthogonal in the infinite volume limit (see Appendix A).
In the following for simplicity
we will use the notation $|0\rangle_{{e,\mu}}\equiv |0(0)\rangle_{{e,\mu}}$ to denote the
flavor vacuum state at the reference time $t=0$.

 The explicit expression of
the flavor annihilation/creation operators for ${\bf
k}=(0,0,|{\bf k}|)$ is:
\begin{equation}\label{BVmatrix}
\left(\begin{array}{c}
{\alpha}^{r}_{{\bf k},\nu_{e}}(t)\\
{\alpha}^{r}_{{\bf k},\nu_{\mu}}(t)\\
{\beta}^{r\dag}_{{-\bf k},\nu_{e}}(t)\\
{\beta}^{r\dag}_{{-\bf k},\nu_{\mu}}(t)
\end{array}\right)
\, = \, \left(\begin{array}{cccc} c_\theta &  s_\theta\, U^{*}_{{\bf
k}}(t) &0 & s_\theta \, V_{{\bf k}}(t)
\\
- s_\theta\, U_{{\bf k}}(t) & c_\theta &  s_\theta \, V_{{\bf k}}(t)
& 0
\\
0& - s_\theta \,V_{{\bf k}}(t) &c_\theta & s_\theta \,U^{*}_{{\bf k}}(t)
\\
- s_\theta \,V_{{\bf k}}(t) & 0 & - s_\theta\, U_{{\bf k}}(t) &
c_\theta
\\
\end{array}\right)
\left(\begin{array}{c}
\alpha^{r}_{{\bf k},1}\\
\alpha^{r}_{{\bf k},2}\\
\beta^{r\dag}_{{-\bf k},1}\\
\beta^{r\dag}_{{-\bf k},2}
\end{array}\right)
\end{equation}
where $c_\theta\equiv \cos\theta$, $s_\theta\equiv \sin\theta$ and
\begin{eqnarray}
&&U_{{\bf k}}(t) \equiv u^{r\dag}_{{\bf k},2}(t) u^{r}_{{\bf k},1}(t) =
v^{r\dag}_{-{\bf k},1}(t) v^{r}_{-{\bf k},2}(t)  \\
\label{2.37} &&V_{{\bf k}}(t) \equiv \epsilon^{r}\;u^{r\dag}_{{\bf
k},1}(t) v^{r}_{-{\bf k},2}(t) = -\epsilon^{r}\;u^{r\dag}_{{\bf
k},2}(t) v^{r}_{-{\bf k},1}(t),
\end{eqnarray}
with $\epsilon^{r}=(-1)^{r}$.
We have:
\bea
U_{{\bf k}}(t)= |U_{{\bf k}}|e^{i(\omega_{k,2}-\omega_{k,1})t}\;\;
,\qquad \qquad V_{{\bf k}}(t)= |V_{{\bf k}}|e^{i(\omega_{k,2}+\omega_{k,1})t}
\eea

\begin{eqnarray}
|U_{{\bf k}}|&=&
\left(\frac{\omega_{k,1}+m_{1}}{2\omega_{k,1}}\right)^{\frac{1}{2}}
\left(\frac{\omega_{k,2}+m_{2}}{2\omega_{k,2}}\right)^{\frac{1}{2}}
\left(1+\frac{|{\bf k}|^{2}}{(\omega_{k,1}+m_{1})
(\omega_{k,2}+m_{2})}\right)\\
\label{2.39} |V_{{\bf k}}|&=&
\left(\frac{\omega_{k,1}+m_{1}}{2\omega_{k,1}}
\right)^{\frac{1}{2}}
\left(\frac{\omega_{k,2}+m_{2}}{2\omega_{k,2}}\right)^{\frac{1}{2}}
\left(\frac{|{\bf k}|}{(\omega_{k,2}+m_{2})}-
\frac{|{\bf k}|}{(\omega_{k,1}+m_{1})}\right),
\end{eqnarray}
\begin{equation}\label{2.40}
|U_{{\bf k}}|^{2}+|V_{{\bf k}}|^{2}=1.
\end{equation}

As discussed in Ref.\cite{currents}, in the two flavor mixing case
the group structure associated with mixing transformations is $SU(2)$ and one can define
the following charges in the mass basis:
\bea
Q_{m,j}(t)\,=\,\frac{1}{2}\int d^{3}{\bf x}\,   \nu_m^\dag(x) \,
\tau_j\, \nu_m(x),\quad \qquad  j \,=\, 1, 2, 3,
\eea
where
$\nu_m^T=(\nu_1,\nu_2)$ and  $\tau_j=\si_j/2$ with $\si_j$ being
the Pauli matrices.
The $U(1)$ Noether charges associated with  $\nu_1$ and $\nu_2$ can be then expressed
as
\bea\label{su2noether} Q_{1}\, \equiv \,\frac{1}{2}Q \,+ \,Q_{m,3}
\qquad; \qquad Q_{2}\, \equiv \,\frac{1}{2}Q \,- \,Q_{m,3}.
\eea
with $Q$ total (conserved) charge.
As usual, we need to normal order such charge operators:
\bea
:\wQ_{i}:\, \equiv \,\int d^{3}{\bf x} \, : \nu_{i}^{\dag}(x)\;\nu_{i}(x):
 = \, \sum_{r}
\int d^3 {\bf k} \, \lf( \al^{r\dag}_{{\bf k},i}
\al^{r}_{{\bf k},i}\, -\, \beta^{r\dag}_{-{\bf
k},i}\beta^{r}_{-{\bf k},i}\ri)\, \,, \qquad i=1,2.
\eea
where the $:..:$ denotes normal ordering with respect to the vacuum $|0\ran_{1,2}$.

It is then clear that the neutrino states with definite masses  defined as
\bea
|\nu^{r}_{\;{\bf k},i}\ran =
\al^{r\dag}_{{\bf k},i} |0\ran_{1,2}, \qquad i=1,2,
\eea
are eigenstates of the above conserved charges, which can be identified with the
lepton charges in the absence of mixing.

The situation changes when we turn to the flavor basis \cite{currents}:
\bea
Q_{f,j}(t)\,=\,\frac{1}{2}\int d^{3}{\bf x}\,   \nu_f^\dag(x) \,
\tau_j\, \nu_f(x),\quad \qquad  f \,=\, e, \mu,\quad \qquad  j = 1,2
\eea
where
$\nu_f^T=(\nu_e,\nu_\mu)$.
Observe that  the diagonal $SU(2)$
generator $Q_{f,3}(t)$ is time-dependent in the flavor basis.

 Thus the flavor charges defined as
\bea Q_{\nu_e}(t)=\frac{1}{2}Q + Q_{f,3}(t)\qquad; \qquad
Q_{\nu_\mu}(t)=\frac{1}{2}Q - Q_{f,3}(t),
\eea
are now time-dependent and are the lepton charges in presence
of mixing \cite{currents}. Indeed their expectation values in the flavor state
(see Eq.(\ref{flavstate}) below) give the oscillation formulas
\cite{BHV98,BCRV01,yBCV02,spaceosc,Blaspalm}.

Particular attention has to be paid now to the normal ordering issue. We define
the normal ordered charges $\nof \wwQ_{\nu_\sigma}(t) \nof\,$ with respect to
the vacuum $|0\ran_{e,\mu}$ as
\bea
\nof \wwQ_{\nu_\sigma}(t) \nof\, \equiv \,\int d^{3}{\bf x}\,
\nof \nu_{\sigma}^{\dag}(x)\;\nu_{\sigma}(x) \nof \,= \, \sum_{r}
\int d^3 {\bf k} \, \lf( \al^{r\dag}_{{\bf k},\nu_{\si}}(t)
\al^{r}_{{\bf k},\nu_{\si}}(t)\, -\, \beta^{r\dag}_{-{\bf
k},\nu_{\si}}(t) \beta^{r}_{-{\bf k},\nu_{\si}}(t)\ri)\, , \qquad
\sigma=e,\mu;,
\eea
where the new symbol $\nof ... \nof\,$ for the normal ordering was introduced to
remember that it refers to the flavor vacuum. The definition for any operator $A$,
is the following
\be \lab{nordf}
 \nof A \nof\, \equiv \,  A \, -\, {}_{e,\mu}\lan 0| A | 0 \ran_{e,\mu}
\ee
Note that:
\bea\lab{Qeq} && \nof \wwQ_{\nu_\sigma}(t) \nof \,= \,G_\theta^{-1}(t)\;  :\wQ_j :\;
G_\theta(t)\; ,\;  \qquad with \qquad(\sigma,j) = (e,1),(\mu,2),
\eea
and
\bea
\nof {\wwQ}_{\nu_e}(t) \nof \,+\, \nof{\wwQ}_{\nu_\mu}(t)\nof\,= \,:{\wQ}_{1}:\,+\,:\wQ_{2}:\,=\,:\wQ:.
\eea

We define the flavor states as eigenstates of the flavor
charges $\wwQ_{\nu_\sigma} $ at a reference  time $t=0$:
\bea\label{flavstate}
|\nu^{r}_{\;{\bf k},\si}\ran \equiv \al^{r\dag}_{{\bf k},{\nu_{\sigma}}}(0) |0(0)\ran_{{e,\mu}},
\qquad \si = e,\mu\eea
 and similar ones for antiparticles.
We have
\bea \label{su2flavstates}
\nof \wwQ_{\nu_e}(0)\nof|\nu^{r}_{\;{\bf k},e}\rangle\,=\,|\nu^{r}_{\;{\bf k},e}\rangle\qquad ;\qquad
\nof\wwQ_{\nu_\mu}(0)\nof|\nu^{r}_{\;{\bf k},\mu}\rangle\,=\,|\nu^{r}_{\;{\bf k},\mu}\rangle \eea
and $\nof\wwQ_{\nu_e}(0)\nof\,|\nu^{r}_{\;{\bf k},\mu}\rangle\,=\,
\nof\wwQ_{\nu_\mu}(0)\nof\,|\nu^{r}_{\;{\bf k},e}\rangle\,=\,0$.
Moreover
\bea \label{Oflavstates}
 \nof\wwQ_{\nu_\sigma}(0)\nof\;|0\rangle_{{e,\mu}}\,=\,0.\eea
These results are far from being trivial since the usual Pontecorvo
states \cite{chengli}:
\begin{eqnarray} \label{nue0a}
|\nu^{r}_{\;{\bf k},e}\rangle_P &=& \cos\theta\;|\nu^{r}_{\;{\bf k},1}\rangle \;+\;
\sin\theta\; |\nu^{r}_{\;{\bf k},2}\rangle \;
\\ [2mm] \label{nue0b}
|\nu^{r}_{\;{\bf k},\mu}\rangle_P &=& -\sin\theta\;|\nu^{r}_{\;{\bf k},1}\rangle \;+\;
\cos\theta\; |\nu^{r}_{\;{\bf k},2}\rangle \; ,
\end{eqnarray}
are {\em not} eigenstates of the flavor charges, as can be easily checked.
%

It is instructive to consider the expectation values of the flavor charges onto the
Pontecorvo states, in order to better appreciate how much the lepton charge is
violated in the usual quantum mechanical states. We find:
\bea\label{Qnu}
\;_{P}\langle\nu^{r}_{\,{\bf k},e}| \nof \wwQ_{\nu_e}(0)\nof |\nu^{r}_{\,{\bf k},e}\rangle_{P}
= \cos ^{4}\theta + \sin^{4}\theta + 2 |U_{\bf k}| \sin^{2}\theta \cos^{2}\theta
+\sum_{r}\int d^{3}{\bf k},
\eea
and
%
\bea\label{Q0}
\;_{1,2}\langle 0 |\nof \wwQ_{\nu_e}(0) \nof| 0 \rangle_{1,2} = \sum_{r}\int d^{3}{\bf k},
\eea
Eqs.(\ref{Qnu}) and (\ref{Q0}) clearly are both infinite.

One may think that the problem with infinity is due to the normal ordering with respect to the flavor vacuum
and consider the expectation values of  $:\wQ_{\nu_\sigma}(t):$, i.e. the
normal ordered  flavor charges with respect to
the mass vacuum $|0\ran_{1,2}$. One has then
\bea
\;_{P}\langle\nu^{r}_{\,{\bf k},e}|: \wQ_{\nu_e}(0): |\nu^{r}_{\,{\bf k},e}\rangle_{P}
= \cos ^{4}\theta + \sin^{4}\theta + 2 |U_{\bf k}| \sin^{2}\theta \cos^{2}\theta < 1,
\quad \forall \theta \neq 0, \quad m_{1} \neq m_{2}, \quad {\bf k} \neq 0,
\eea
\bea
\;_{1,2}\langle 0 |: \wQ_{\nu_e}(0): | 0 \rangle_{1,2} = 0,
\eea
and
\bea
\;_{1,2}\langle 0 | (:\wQ_{\nu_e}(0):)^{2} | 0 \rangle_{1,2} =
4\sin^{2}\theta \cos^{2}\theta \int d^{3}{\bf k}|V_{\bf k}|^{2},
\eea
\bea
\;_{P}\langle\nu^{r}_{\,{\bf k},e}| (:\wQ_{\nu_e}(0):)^{2} | \nu^{r}_{\,{\bf k},e}\rangle_{P}=
\cos^{6}\theta + \sin^{6}\theta + \sin^{2}\theta \cos^{2}\theta
\lf[2 |U_{\bf k}| + |U_{\bf k}|^{2} + 4 \int d^{3}{\bf k}|V_{\bf k}|^{2} \ri],
\eea
which are both infinite, thus making the corresponding quantum fluctuations divergent.

Hence, we conclude that the correct flavor state and normal ordered operators are those
defined in
Eqs.(\ref{flavstate}) and (\ref{nordf}) respectively.

\section{Mass parameters and field mixing}

In Eq.(\ref{exnue1}) $u^{r}_{{\bf k}, i}$ and $v^{r}_{-{\bf k},
i}$ are the spinor wavefunctions of the massive neutrinos
$\nu_{i}, ~i = 1,2$.  As already observed in the previous section,
Eq.(\ref{exnue1}) provides an expansion of the flavor fields
$\nu_{\sigma}, ~\sigma =e,\mu$, in the same basis of $\nu_{i}, ~i
= 1,2$. However, it was noticed in Ref. \cite{Fujii:1999xa} that
expanding the flavor fields in the same basis as the (free) fields
with definite masses is actually a special choice, and that in principle a more
general possibility exists. Indeed, in the expansion
(\ref{exnue1}) one could use eigenfunctions with arbitrary
masses $\mu_\sigma$. In other words, the transformation
(\ref{BVoper}) can be generalized by writing the flavor fields as
\cite{Fujii:1999xa}
\begin{eqnarray}\label{exnuf2}
\nu_{\sigma}(x) &=& \frac{1}{\sqrt{V}} \sum_{{\bf k},r}
  \left[ u^{r}_{{\bf k},\sigma}(t)
{\widetilde \alpha}^{r}_{{\bf k},\nu_{\sigma}}(t) + v^{r}_{-{\bf
k},\sigma}(t)  {\widetilde \beta}^{r\dag}_{-{\bf k},\nu_{\sigma}}(t) \right]
e^{i {\bf k}\cdot{\bf x}} ,
\end{eqnarray}
where $u_{\sigma}$ and $v_{\sigma}$ are the helicity
eigenfunctions with mass $\mu_\sigma$ (the use of such helicity
eigenfunctions as a basis simplifies calculation with respect to
the choice of Ref.\cite{BV95}). In Eq.(\ref{exnuf2}) the generalized
flavor operators are denoted by a tilde in order to distinguish
them from the ones defined in Eq.(\ref{BVoper}). The expansion
Eq.(\ref{exnuf2}) is more general than the one in
Eq.(\ref{exnue1}) since the latter corresponds to the particular
choice $\mu_e\equiv m_1$, $\mu_\mu \equiv m_2$.

Since the issue of the arbitrary mass parameters in the field
mixing formalism has attracted some attention
\cite{Fujii:1999xa,BCRV01,JM011,yBCV02,Giunti:2003dg}, it is worth
clarifying some basic facts about the choice of
mass parameters within QFT in general, independently from the
occurrence of the field mixing phenomenon.

We will refer to fermion fields since in this paper we are
interested in neutrinos, but the conclusions can be also extended
to boson fields.

First of all, it is worth noting that the mass parametrization problem can be revealed also
in the free field case.
Indeed, one may still consider the change of
mass parametrization $m \rightarrow \mu $, which correspond to choosing ${\widetilde u}^{r}_{{\bf k}}$,
 ${\widetilde v}^{r}_{{\bf k}}$ as free field amplitudes with the arbitrary mass parameter $\mu$.

Consider the set of free (fermion) field operators composed, for
simplicity, by only two elements, i.e. assume our operators are
\begin{equation}\label{FHYoper1}
 \left(\begin{array}{c}
\alpha^{r}_{{\bf k},i}\\
\beta^{r\dag}_{{-\bf k},i}
\end{array}\right) ~, ~~\mathrm{with}~~i = 1,2.
\end{equation}

Let the non-zero masses be $m_{1}$ and $m_{2}$, with $m_{1} \neq
m_{2}$. The wave functions  $u_{i}$ and $v_{i}$ (we omit the
index $\bf k$ whenever no confusion arises) satisfy the free Dirac
equations
\begin{equation}\label{d1i}
 (i \!\not\!k + m_{i})u_{i}  =  0~~,
~~~~(i \!\not\!k - m_{i})v_{i}  =  0 ~~~,
\end{equation}
respectively. Let $|0\rangle_{1,2}$ be the vacuum state
annihilated by $\alpha^{r}_{{\bf k},i}$ and $\beta^{r}_{{-\bf
k},i}$.

Since in QFT there exist infinitely many unitarily inequivalent
representations of the canonical (anti-)commutation relations
\cite{Itz,Um1}, one could consider another
fermion set of operators, say
\begin{equation}\label{FHYoper2}
\left(\begin{array}{c}
{\widetilde{\alpha}}^{r}_{{\bf k},i}(t)\\
{\widetilde{\beta}}^{r\dag}_{{-\bf k},i}(t)
\end{array}\right) ~, ~~ i =1,2,
\end{equation}
related to Eq.(\ref{FHYoper1}) by a Bogoliubov transformation (see below). The freedom of
choosing another set of operators is, for example, typically
exploited in QFT at finite temperature, or more generally when one
introduces the irreducible set of "bare" field operators in terms
of which the Lagrangian of the theory is written. In such a case
the set of bare fields is not necessarily composed by the same
number of elements as the one of the set of physical (asymptotic)
fields satisfying free field equations and in terms of which
observables are expressed. In general, indeed, bound states of
bare fields may also belong to the set of physical fields. The
mapping between the bare fields and the asymptotic fields is
called the Haag expansion \cite{Itz,Um1}.

Suppose the wave functions of the field operators in Eq.(\ref{FHYoper2})
also satisfy the free Dirac equations
\begin{equation}\label{d1}
 (i \!\not\!k + \mu_{i})\widetilde{u}_{i}  =  0~~,
~~~~(i \!\not\!k - \mu_{i})\widetilde{v}_{i}  =  0 ~~~,
\end{equation}
respectively. The mass parameter $\mu_{i}$ in Eqs.(\ref{d1})
represents now the mass of the corresponding arbitrarily chosen
fields in Eq.(\ref{FHYoper2}) and therefore it represents an arbitrary
parameter. Our two set of operators are related by the
transformation
\begin{equation}\label{FHYoper3}
\left(\begin{array}{c}
\widetilde{\alpha}^{r}_{{\bf k},i}(t)\\
\widetilde{\beta}^{r\dag}_{{-\bf k},i}(t)
\end{array}\right)
= I^{-1}_{\mu}(t) \left(\begin{array}{c}
\alpha^{r}_{{\bf k},i}\\
\beta^{r\dag}_{{-\bf k},i}
\end{array}\right)
I_{\mu}(t) ~~,
\end{equation}
with
\begin{equation}
\label{Igen1} I_{\mu}(t)= \prod_{{\bf k}, r}\, \exp\left\{ i
\mathop{\sum_{i}} \xi_{i}^{\bf k}\left[
\alpha^{r\dag}_{{\bf k},i}\beta^{r\dag}_{{-\bf k},i}e^{2i\omega_{i} t} +
\beta^{r}_{{-\bf k},i}\alpha^{r}_{{\bf k},i}e^{-2i\omega_{i} t} \right]\right\}
\end{equation}
where $\xi_{i}^{\bf k}\equiv (\widetilde{\chi}_i - \chi_i)/2$ and
$\cot\widetilde{\chi}_i = |{\bf k}|/\mu_i$, $\cot\chi_i = |{\bf
k}|/m_i$. Notice that for $\mu_1\equiv m_1$, $\mu_2 \equiv m_2$
one has $I_{\mu}(t)=1$, as it must be for the identity
transformation.

The explicit  matrix form of Eq.(\ref{FHYoper3}), written for both
$i$ values, is:
\begin{equation}\label{FHYmatrix1}
\left(\begin{array}{c}
\widetilde{\alpha}^{r}_{{\bf k},1}(t)\\
\widetilde{\alpha}^{r}_{{\bf k},2}(t)\\
\widetilde{\beta}^{r\dag}_{{-\bf k},1}(t)\\
\widetilde{\beta}^{r\dag}_{{-\bf k},2}(t)
\end{array}\right)
\, = \, \left(\begin{array}{cccc}  \rho^{\bf k *}_{1}(t) & 0 &i
\,\lambda^{\bf k}_{1}(t) & 0
\\
0 & \rho^{\bf k *}_{2 }(t) &0 & i  \,\lambda^{\bf k}_{ 2}(t)
\\
i  \,\lambda^{\bf k *}_{1}(t) & 0 & \rho^{\bf k}_{1}(t) & 0
\\
0 & i  \lambda^{\bf k *}_{2}(t) & 0 &  \rho^{\bf k}_{ 2}(t)
\\
\end{array}\right)
\left(\begin{array}{c}
\alpha^{r}_{{\bf k},1}\\
\alpha^{r}_{{\bf k},2}\\
\beta^{r\dag}_{{-\bf k},1}\\
\beta^{r\dag}_{{-\bf k},2}
\end{array}\right)
\end{equation}
where
\begin{eqnarray}\label{rho}
 \rho^{\bf k *}_{ i}(t) \delta_{rs}
= \widetilde{u}^{r\dag}_{{\bf k},i}(t) u^{s}_{{\bf k},i}(t) =
\widetilde{v}^{r\dag}_{-{\bf k},i}(t) v^{s}_{-{\bf k},i}(t)
&\equiv& e^{i(\widetilde{\omega}_{k,i}-\omega_{k,i})} \cos \xi_{i}^{\bf k}\delta_{rs}
\\ \label{lambda}
i \; \lambda^{\bf k *}_{i}(t)\delta_{rs}  = \widetilde{u}^{r\dag}_{{\bf k},i}(t)
v^{s}_{-{\bf k},i}(t) = \widetilde{v}^{r\dag}_{-{\bf k},i}(t) u^{s}_{{\bf k},i}(t)
&\equiv& i\; e^{-i(\widetilde{\omega}_{k,i}+\omega_{k,i})}\sin
\xi_{i}^{\bf k}  \delta_{rs}
\end{eqnarray}
with $i = 1, 2$ and $\widetilde{\omega}_{k,i} = \sqrt{{\bf k}^{2}+\mu_{i}^{2}} $. The vacuum state annihilated by the
($\widetilde{\alpha}^{r}_{{\bf k},i}$, $\widetilde{\beta}^{r}_{-{\bf k},i}$)
operators is
\begin{equation}\label{FHYvac1}
|{\widetilde{0}(t)}\rangle_{1,2}\equiv I^{-1}_{\mu}(t)|0\rangle_{1,2} ~~.
\end{equation}
We observe that Eq.(\ref{FHYoper3}) is indeed nothing but the
Bogoliubov transformation which relates the field operators
($\alpha^{r}_{{\bf k},i}$, $\beta^{r}_{-{\bf k},i}$)  and
($\widetilde{\alpha}^{r}_{{\bf k},i}$,
$\widetilde{\beta}^{r}_{{-\bf k},i}$), of masses $m_i$ and $\mu_{i}$,
respectively. In the infinite volume limit, the Hilbert spaces
where the operators $\alpha^{r}_{{\bf k},i}$ and
$\widetilde{\alpha}^{r}_{{\bf k},i}$ are respectively
defined turn out to be unitarily inequivalent spaces. Moreover,
the transformation parameter $\xi^{\bf k}_{i}$ acts as a label
specifying Hilbert spaces unitarily inequivalent among themselves
for  each (different) value of the $\mu_{i}$ mass parameter.

We note that the vacuum $|{\widetilde{0}(t)}\rangle_{1,2}$ is not
annihilated by $\alpha^{r}_{{\bf k},i}$ and $\beta^{r}_{{\bf k},i}$
and it is not eigenstate
of the the number operators $N_{\alpha_i} = \sum_{r}\int d^{3}{\bf k}\;{\alpha}^{r \dag}_{{\bf k},i}
\alpha^{r}_{{\bf k},i}$ and $N_{\beta_i} = \sum_{r}\int d^{3}{\bf k}\;{\beta}^{r \dag}_{{\bf k},i}
\beta^{r}_{{\bf k},i}$.
Similarly $|0 \rangle_{1,2}$ is not annihilated by $\widetilde{\alpha}^{r}_{{\bf k},i}$
and $\widetilde{\beta}^{r}_{{\bf k},i}$ and it is not eigenstate of the the number
operators $\widetilde{N}_{\alpha_{i}}(t) = \sum_{r}\int d^{3}{\bf k}\;\widetilde{\alpha}^{r \dag}_{{\bf k},i}(t)
\widetilde{\alpha}^{r}_{{\bf k},i}(t)$ and
$\widetilde{N}_{\beta_{i}}(t) = \sum_{r}\int d^{3}{\bf k}\;\widetilde{\beta}^{r \dag}_{{\bf k},i}(t)
\widetilde{\beta}^{r}_{{\bf k},i}(t)$. One obtains
\begin{equation}\label{vac1}
{}_{1,2}\langle{\widetilde{0}(t)}|N_{\alpha_i}|{\widetilde{0}(t)}\rangle_{1,2} =
{}_{1,2}\langle{\widetilde{0}(t)}|N_{\beta_i}|{\widetilde{0}(t)}\rangle_{1,2} =
{\sin}^{2} \xi^{\bf k}_{i}~,
\end{equation}
and
\begin{equation}\label{vac2}
{}_{1,2}\langle{0}|\widetilde{N}_{\alpha_i}(t)|{0}\rangle_{1,2} =
{}_{1,2}\langle{0}|\widetilde{N}_{\beta_i}(t)|{0}\rangle_{1,2} = {\sin}^{2} \xi^{\bf k}_{i}~.
\end{equation}

In other words, the number operator, say $N_{\alpha_i}$,
is not an invariant quantity under
the Bogoliubov transformation Eq.(\ref{FHYoper3}): it gets a
dependence on the mass parameters.
 This is, however, not surprising
since it is known that the Bogoliubov transformation
Eq.(\ref{FHYoper3}) introduces a {\it new} set of canonical operators
and a {\it new} (i.e. in the infinite volume limit unitarily, and
therefore physically, inequivalent) Hilbert space. Stated
differently, through the Bogoliubov transformation a {\it new} set
of asymptotic fields (a {\it new} set of quasiparticles) is
introduced, i.e. there are infinitely many sets of asymptotic
fields, each set being associated to its specific representation.
The choice of which one is the set to be used is then
dictated by the physical conditions which are actually realized.
For example, the mass values which have to be singled out in the
renormalization procedure must be the observed physical masses.

Since the tilde quantities correspond to
some new quasi-particle objects and the tilde number operator
describes a different type of particles,
then, the number operator average shall not be expected to remain the
same under such transformations. Indeed, defined the state
$|{\widetilde \alpha}^{r}_{{\bf k},i}\rangle$ as
$|{\widetilde \alpha}^{r}_{{\bf k},i}(0)\rangle =
{\widetilde \alpha}^{r \dag}_{{\bf k},i}(0)|{\widetilde 0}\rangle$,
 we have:
\bea
\langle{\widetilde \alpha}^{r}_{{\bf k},i}
(0)\mid{\widetilde N_{\alpha_{i}}(t)} \mid {\widetilde \alpha}^{r}_{{\bf k},i}
(0)\rangle=|\{ {\widetilde \alpha}^{r}_{{\bf k},i}
(t),{\widetilde \alpha}^{r \dag}_{{\bf k},i}
(0)\}|^2=||\rho^{\bf k}_{i}|^2 e^{i({\widetilde \omega}_{k,i}-\omega
  _{k,i})t} +
|\lambda^{\bf k}_{i}|^2 e^{i({\widetilde \omega}_{k,i}+\omega
  _{k,i})t} |^2, \qquad i=1,2,
\eea
which shows that the expectation value of the time dependent
number operator is not preserved by the transformation
(\ref{FHYoper3}) applied to both states and operators.
Nevertheless, in the cases of free fields, the charge operator is
still conserved in transformation
like Eq.(\ref{FHYoper3}):
\bea
{\widetilde Q_{i}} & = & Q_{i},
\eea
moreover the expectation value of the charge at time $t$
on the state at time $t=0$ is also free from mass parameters
\bea
\langle {\widetilde \alpha}^{r}_{{\bf k},{i}}
(0)\mid{\widetilde Q_{i}(t)} \mid {\widetilde \alpha}^{r}_{{\bf k},{i}}
(0)\rangle & = & \langle {\alpha}^{r}_{{\bf k},{i}}
(0)\mid{Q_{i}(t)} \mid { \alpha}^{r}_{{\bf k},{i}}(0)\rangle,
\eea
since we have
\bea
|\{ {\widetilde \alpha}^{r}_{{\bf k},{i}}
(t),{\widetilde \alpha}^{r \dag}_{{\bf k},{i}}
(0)\}|^2 + |\{ {\widetilde \beta}^{r \dag}_{{\bf k},{i}}
(t),{\widetilde \alpha}^{r \dag}_{{\bf k},{i}}
(0)\}|^2 = |\{ \alpha^{r}_{{\bf k}_{i}}
(t),\alpha^{r \dag}_{{\bf k},{i}}
(0)\}|^2 + |\{ \beta^{r \dag}_{{\bf k},{i}}
(t),\alpha^{r \dag}_{{\bf k},{i}}
(0)\}|^2.
\eea

Similar results are obtained in the
two flavor particle mixing case.
Note, however, that in the case of three flavor mixing and in presence of CP violation,
the charge operator and the flavor states are dependent on the arbitrary mass parameters
and the quantity which is invariant under the transformation like the Eq.(\ref{FHYoper3})
is
\bea
\langle \alpha^{r}_{{\bf k, \nu_\sigma}}|\nof\wwQ_{\nu_{\sigma}}(t)\nof|\alpha^{r}_{{\bf k, \nu_\sigma}}\rangle
-{}_{e,\mu,\tau}\langle{0}|\nof\wwQ_{\nu_{\sigma}}(t)\nof|{0}\rangle_{e,\mu,\tau},
\eea
with $|\alpha^{r}_{{\bf k, {\nu_{\sigma}}}}\rangle =
\alpha^{r \dag}_{{\bf k, {\nu_{\sigma}}}}(0)|0\rangle_{e,\mu,\tau}$
and $\sigma = e, \mu, \tau$ as observed in Ref.\cite{yBCV02}

Having clarified that the possibility of different mass parameters is
intrinsic to the very same structure of QFT and is independent of
the occurrence or not of the field mixing, we may affirm that
the mass parameters must be chosen not arbitrarily,
but they must be justified on the ground of physical reasons \cite{Blasone:1999jb,JM011,yBCV02}.

In particular, in the mixing problem, the choice $\mu_e \equiv m_1$, $\mu_\mu \equiv m_2$,
$\mu_\tau \equiv m_3$, is
motivated by the fact that $m_1$, $m_2$ and $m_3$ are the masses
of the mass eigenfields and therefore such a choice is the only
one physically relevant.

In our computations, instead of using the number operator, we use the charge operator
$Q$ which in the mixing phenomena describes
the relative population densities of flavor particles in the beam and it is related
with the oscillating observables that are: the lepton charge, in the case of neutrino mixing,
the strange charge in meson systems like $K^{0}-\bar{K}^{0}$ and $B_{s}^{0}-\bar{B}_{s}^{0}$,
the charmed charge in the systems $D^{0}-\bar{D}^{0}$, and so on.
We note that, as shown in refs.\cite{Blaspalm,Capolupo:2004pt}, the momentum operator for
the mixing of neutral fields plays an analogous role to the one of the charge operator
for charged fields.

\section{Expectation value of neutrino lepton charge in the
one pion state in pion decay}

In this Section we study the structure of the flavor charge expectation values in
the case the production process of neutrinos is taken into account. This is done by
using the flavor Hilbert space discussed above.
In Ref.\cite{Fujii:2004yd} a similar calculation was performed by using the mass Hilbert space,
so neglecting flavor vacuum effects.

The final aim of the authors of Ref.\cite{Fujii:2004yd} was to derive an oscillation formula in space, which
is relevant for current experiments. On the other hand, a general oscillation formula with space-time
dependence has been obtained in Ref.\cite{spaceosc} in terms of expectation values of the flavor currents on
the flavor neutrino states, exhibiting the corrections due to the flavor vacuum.
In the following, we show in an explicit way how calculations can be performed with interacting fields
on the flavor Hilbert space.

We consider the case where the flavor neutrinos are produced
through the pion decay $\pi^{+}\longrightarrow \mu^{+} +
\nu_{\mu}$. We use the phenomenological approach to the pion
decay \cite{Lurie} without referring to the quark structure of the
pion. As initial $\pi^{+}$ state at time $x^{0}_{I} $, we use $|\Pi(\langle{\bf
k}\rangle, {\bf x}, x^{0}_{I})\rangle \equiv |\pi(\langle{\bf
k}\rangle, {\bf x}, x^{0}_{I})\rangle \times
|0(x^{0}_{I})\rangle_{\nu} \times |0\rangle_{\mu}$  with
$\langle{\bf k}\rangle$ the average $k$ vector (see below). The
neutrino vacuum is the flavor vacuum: $|0(x^{0}_{I})\rangle_{\nu} ~=~
G^{-1}_{\theta}(x^{0}_{I})|0\rangle_{1,2}$ (note the change of notation
for $|0(x^{0}_{I})\rangle_{e,\mu}$ with respect to the previous sections).
We calculate the
expectation values of neutrino flavor charge
$\nof \wwQ_{\nu_{\mu}}(x^{0})\nof$ with
respect to one pion state $|\Pi(\langle{\bf k}\rangle, {\bf x},
x^{0})\rangle $ where $x^{0}_{I} < x^{0}$
\bea \nonumber \langle\nof \wwQ_{\nu_{\mu}}(x^{0})\nof\rangle &=& \langle
\Pi(\langle{\bf k}\rangle, {\bf x}, x^{0})
|\nof\wwQ_{\nu_{\mu}}(x^{0})\nof|\Pi(\langle{\bf
k}\rangle, {\bf x}, x^{0})\rangle \label{Q_nu_PI} \\
 &=& \langle \Pi(\langle{\bf
k}\rangle, {\bf x}, x^{0}_{I}) |S^{-1}(x^{0},x^{0}_{I})
\nof\wwQ_{\nu_{\mu}}(x^{0})\nof S(x^{0},x^{0}_{I}) |\Pi(\langle{\bf k}\rangle, {\bf
x}, x^{0}_{I})\rangle \eea
with
\bea |\Pi(\langle{\bf k}\rangle, {\bf x}, x^{0})\rangle =
S(x^{0},x^{0}_{I}) |\Pi(\langle{\bf k}\rangle, {\bf x},
x^{0}_{I})\rangle
\eea
where, in the interaction representation,
the normal ordered neutrino flavor charge
$ \nof\wwQ_{\nu_{\mu}}(x^{0})\nof$  is
\bea \label{3}
\nof\wwQ_{\nu_{\mu}}(x^{0})\nof=\sum_{r}\int d^{3}{\bf
k}\left[\alpha _{{\bf k},\nu_{\mu} }^{r\dagger }(x^{0})\alpha _{{\bf
k},\nu_{\mu} }^{r}(x^{0})-\beta _{-{\bf k} ,\nu_{\mu} }^{r\dagger
}(x^{0})\beta _{-{\bf k},\nu_{\mu} }^{r}(x^{0})\right] \eea
and $S(x^{0},x^{0}_{I})$
is
\bea S(x^{0},x^{0}_{I})=
\sum_{m=0}(-i)^{m}\int^{x^{0}}_{x^{0}_{I}}d^{4}y_{1}
\int^{y_{1}^{0}}_{x^{0}_{I}}
d^{4}y_{2}....\int^{y_{m-1}^{0}}_{x^{0}_{I}}d^{4}y_{m}
H_{\mathrm{int}}(y_{1})
H_{\mathrm{int}}(y_{2})....H_{\mathrm{int}}(y_{m}).
\eea
The muon neutrino field and the muon field are expanded,
in the interaction representation, as
\bea \label{8} \nu_\mu(x)= \sum_{r} \int \frac{d^3 {\bf
q}}{(2\pi)^{\frac{3}{2}}} \lf[ u^{r}_{{\bf q},2}(x^{0})
\al^{r}_{{\bf q},\nu_{\mu}}(x^{0})  e^{i {\bf q}\cdot{\bf x}} +
v^{r}_{{\bf q},2}(x^{0}) \bt^{r\dag}_{{\bf q},\nu_{\mu}}(x^{0})
e^{-i {\bf q}\cdot{\bf x}}\ri] , \eea
\bea\mu(x)= \sum_{s} \int
\frac{d^3 {\bf p}}{(2\pi)^{\frac{3}{2}}} \lf[ u^{s}_{{\bf
p},\mu}(x^{0}) \al^{s}_{{\bf p},\mu} e^{i {\bf p}\cdot{\bf
x}} + v^{s}_{{\bf p},\mu}(x^{0}) \bt^{s\dag}_{{\bf
p},\mu}e^{-i {\bf p}\cdot{\bf x}} \ri]  ,
\eea
with $\al^{r}_{{\bf q},\nu_{\mu}}(x^{0})$ flavor annihilation
operator for neutrino. In Eq.(\ref{8}), $u^{r}_{{\bf q},2}(x^{0})$
is the spinor wave function for the massive neutrino field
$\nu_{2}(x)$.

The hamiltonian of weak interaction is \bea H_{\mathrm{int}}(x)=
-i\;g_{\pi}\;\overline{\nu}_{\mu}(x)\;
\gamma^{\lambda}(1+\gamma^{5})\;\mu(x)\;\partial_{\lambda}\pi(x) +
\mathrm{h.c.,} \eea where $\pi(x)$, $\mu(x)$, $\nu_{\mu}(x)$, are
the fields of pion, muon and flavor (muon) neutrino, respectively.

 In the lowest order (i.e. the second order) of the weak interaction,
 we have
\bea \langle \nof\wwQ_{\nu_{\mu}}(x^{0})\nof\rangle= \langle \Pi(\langle{\bf
k}\rangle, {\bf x}, x^{0}_{I})
|\left[\int^{x^{0}}_{x^{0}_{I}}d^{4}z H_{\mathrm{int}}(z)
\nof \wwQ_{\nu_{\mu}}(x^{0})\nof\int^{x^{0}}_{x^{0}_{I}}d^{4}y H_{\mathrm{int}}(y)
\right]|\Pi(\langle{\bf k}\rangle, {\bf x}, x^{0}_{I})\rangle \eea
and, in the case of positively charged pion $\pi^{+}$, we
have explicitly
\bea \langle \nof\wwQ_{\nu_{\mu}}(x^{0})\nof\rangle &=&\int^{x^{0}}_{x^{0}_{I}}
\int^{x^{0}}_{x^{0}_{I}}d^{4}z d^{4}y
\langle \pi(\langle{\bf k}\rangle, {\bf x}, x^{0}_{I}) |\\
\non &\times&  \Big[i
g^{*}_{\pi}\partial_{\lambda}\pi^{\dag}(z)\;_{\mu}\langle 0|
\overline{\mu}(z)\gamma^{\lambda}(1+\gamma_{5}) \;_{\nu}\langle
0(x^{0}_{I})|{\nu}_{\mu}(z) \nof\wwQ_{\nu_{\mu}}(x^{0})\nof
\overline{\nu}_{\mu}(y)|0(x^{0}_{I})\rangle_{\nu}
\\\non &\times & \gamma^{\sigma}(1+\gamma_{5}) \mu(y)
|0\rangle_{\mu}\;i g_{\pi }
\partial_{\sigma}\pi(y)\Big]
|\pi(\langle{\bf k}\rangle, {\bf x}, x^{0}_{I})\rangle.
\eea

Expressing the charge operator as in Eq.(\ref{3}), a
straightforward calculation gives
\bea\label{CHARGE} \non
\langle \nof\wwQ_{\nu_{\mu}}(x^{0})\nof \rangle &=&
-|g_{\pi}|^{2} \int^{x^{0}}_{x^{0}_{I}}
\int^{x^{0}}_{x^{0}_{I}}d^{4}z d^{4}y \langle \pi(\langle{\bf
k}\rangle, {\bf x}, x^{0}_{I})|
\partial_{\lambda}\pi^{\dag}(z)
\partial_{\sigma}\pi(y)| \pi(\langle{\bf k}\rangle, {\bf x}, x^{0}_{I})
\rangle\;
\\\non&\times &\sum_{r,t,u,v}\int \frac{d^{3}{\bf p}\;
d^{3}{\bf q}\; d^{3}{\bf q'}\; d^{3}{\bf q''}}{(2\pi)^{6}}\;
\Big(e^{i({\bf p}+{\bf q})\cdot{\bf z}} e^{-i({\bf p}+{\bf
q'})\cdot {\bf y}}
\\\non &\times &
e^{-i(\omega_{p}+\omega_{q,2})z^{0}}
e^{i(\omega_{p}+\omega_{q',2})y^{0}} \bar{v}^{r}_{{\bf
p},\mu}\gamma^{\lambda}(1+\gamma_{5}) u^{t}_{{\bf
q},2}\bar{u}^{u}_{{\bf q'},2}\gamma^{\sigma}(1+\gamma_{5})
v^{r}_{{\bf p},\mu}
\\\non &\times & \;_{\nu}\langle 0(x^{0}_{I})
|\al^{t}_{{\bf q},\nu_{\mu}}(z^{0})[\al^{v \dag}_{{\bf
q''},\nu_{\mu}}(x^{0}) \al^{v }_{{\bf
q''},\nu_{\mu}}(x^{0})-\beta^{v \dag}_{{\bf q''},\nu_{\mu}}(x^{0})
\beta^{v }_{{\bf q''},\nu_{\mu}}(x^{0})]\al^{u \dag}_{{\bf
q'},\nu_{\mu}}(y^{0}) |0(x^{0}_{I})\rangle_{\nu}
\\ &+& similar \; terms, \eea
being $\omega_{p}=\sqrt{{\bf p}^{2}+m^{2}_{\mu}}$ and $\omega_{q,2}=\sqrt{{\bf q}^{2}+m^{2}_{2}}$.

The explicit expression of Eq.(\ref{CHARGE}) is given by Eq.(\ref{Echarge}) in Appendix B.

\vspace{5mm}
\noindent \emph{Evaluation of the Bosonic term}
\vspace{5mm}

We consider first the bosonic part: $\langle \pi(\langle{\bf
k}\rangle, {\bf x}, x^{0}_{I})|
\partial_{\lambda}\pi^{\dag}(z)
\partial_{\sigma}\pi(y)| \pi(\langle{\bf k}\rangle, {\bf x}, x^{0}_{I})
\rangle\;.$
The pion state is defined as:
\bea | \pi(\langle{\bf k}\rangle, {\bf x},
x^{0}_{I})\rangle\;=\int d^{3}k\; \frac{A_{\pi}({\bf
k},\langle{\bf k}\rangle) \;e^{-i{\bf k} \cdot{\bf x}+i
\Omega_{ k}x^{0}_{I}}} {\sqrt{2 \Omega_{k}}}\;a_{{\bf
k}\pi}^{\dagger} | 0\rangle_{\pi}
\eea
with \bea A_{\pi}({\bf
k},\langle{\bf k}\rangle)\; = \frac{1}{(\sqrt{2
\pi}\sigma^{2}_{\pi})^{3/2}}\; \exp\Big[-\frac{\lf({\bf
k}-\langle{\bf k}\rangle\ri)^{2}} {4 \sigma^{2}_{\pi}}\Big],
\eea
and $\Omega_{k}=\sqrt{{\bf k}^{2}+m^{2}_{\pi}}$.
The pion field in the interaction picture is:
\bea \pi(x)= \int \frac{d^3{\bf k}}{(2\pi)^{\frac{3}{2}}}
\frac{1}{\sqrt{2\Omega_{k}}} \lf( a_{\bf k}\ e^{i({\bf k \cdot x} -
\Omega_{k}t)} + b^{\dag}_{\bf k} \ e^{-i({\bf k \cdot x}-\Omega_{k} t)}\ri),
\eea
hence
\bea \non &&
 \langle \pi(\langle{\bf k}\rangle, {\bf x}, x^{0}_{I})
|\partial_{\lambda}\pi^{\dag}(z)
\partial_{\sigma}\pi(y)|
\pi(\langle{\bf k}\rangle, {\bf x}, x^{0}_{I})\rangle\;=
\\ &&=
  \int\frac{d^3{\bf k}d^3{\bf k'}}{4(2\pi)^{3}
{\Omega_{ k}\Omega_{ k'}}} A^{\ast}_{\pi}({\bf
k},\langle{\bf k}\rangle) A_{\pi}({\bf k'},\langle{\bf
k}\rangle)k_{\lambda}k'_{\sigma} e^{-i {\bf k \cdot z}}e^{i
{\bf k' \cdot y}}
 e^{i \Omega_{k}(z^{0}-x^{0}_{I})} e^{-i
\Omega_{k'}(y^{0}-x^{0}_{I})}e^{i ({\bf k}-{\bf k'}) \cdot {\bf x}}, \eea
\vspace{5mm} thus Eq.(\ref{CHARGE}) may be expressed as

\bea\label{CHARGE1} \non \langle \nof\wwQ_{\nu_{\mu}}(x^{0})\nof\rangle &=&
-|g_{\pi}|^{2} \int^{x^{0}}_{x^{0}_{I}}
\int^{x^{0}}_{x^{0}_{I}}d^{4}z d^{4}y \sum_{r,t,u,v}\int
\frac{d^3{\bf k}\; d^3{\bf k'} d^{3}{\bf p}\; d^{3}{\bf q}\;
d^{3}{\bf q'}\; d^{3}{\bf q''}}{4(2\pi)^{9}{\Omega_{ k}\Omega_{
k'}}}\; A^{\ast}_{\pi}({\bf k},\langle{\bf k}\rangle)
A_{\pi}({\bf k'},\langle{\bf k}\rangle)
\\\non&\times & k_{\lambda}k'_{\sigma}
e^{i ({\bf k}-{\bf k'}) \cdot {\bf x}}  e^{-i
(\Omega_{k}-\Omega_{k'})x^{0}_{I}}
\\\non&\times &
e^{-i({\bf k}-{\bf p}-{\bf q})\cdot{\bf z}} e^{i({\bf
k'}-{\bf p}-{\bf q'})\cdot {\bf y}}
e^{i(\Omega_{ k}-\omega_{p}-\omega_{q,2})(z^{0}-x^{0}_{I})}
e^{-i(\Omega_{ k'}-\omega_{p}-\omega_{q',2})(y^{0}-x^{0}_{I})}
\bar{v}^{r}_{{\bf
p},\mu}\gamma^{\lambda}(1+\gamma_{5}) u^{t}_{{\bf
q},2}\bar{u}^{u}_{{\bf q'},2}
\\\non &\times & \gamma^{\sigma}(1+\gamma_{5})v^{r}_{{\bf p},\mu}
\;_{\nu}\langle 0(x^{0}_{I})
|\al^{t}_{{\bf q},\nu_{\mu}}(z^{0})[\al^{v \dag}_{{\bf
q''},\nu_{\mu}}(x^{0}) \al^{v }_{{\bf
q''},\nu_{\mu}}(x^{0})-\beta^{v \dag}_{{\bf q''},\nu_{\mu}}(x^{0})
\beta^{v }_{{\bf q''},\nu_{\mu}}(x^{0})]\al^{u \dag}_{{\bf
q'},\nu_{\mu}}(y^{0}) |0(x^{0}_{I})\rangle_{\nu}
\\ &+& similar \; terms, \eea
(see Eq.(\ref{Echarge1}) in Appendix B) and then

\bea\label{CHARGE2} \non \langle \nof\wwQ_{\nu_{\mu}}(x^{0})\nof\rangle &=&
-|g_{\pi}|^{2} \int^{x^{0}}_{x^{0}_{I}}
\int^{x^{0}}_{x^{0}_{I}}dz^{0} dy^{0} \sum_{r,t,u,v}\int
\frac{ d^{3}{\bf k}\;
d^{3}{\bf k'}\;d^{3}{\bf p}\;  d^{3}{\bf q}}{4(2\pi)^{3}{\Omega_{k}\Omega_{
k'}}}\;
\; A^{\ast}_{\pi}({\bf k},\langle{\bf k}\rangle)
A_{\pi}({\bf k'},\langle{\bf k}\rangle)\;
e^{i ({\bf k}-{\bf k'}) \cdot {\bf x}}  e^{-i
(\Omega_{k}-\Omega_{k'})x^{0}_{I}}
\\\non&\times &
e^{i(\Omega_{k}-\omega_{p}-\omega_{k-p,2})(z^{0}-x^{0}_{I})}
e^{-i(\Omega_{k'}-\omega_{p}-\omega_{k'-p,2})(y^{0}-x^{0}_{I})}
\bar{v}^{r}_{{\bf
p},\mu}k_{\lambda}\gamma^{\lambda}(1+\gamma_{5}) u^{t}_{{\bf k}-{\bf p},2}
\bar{u}^{u}_{{\bf k'}-{\bf p},2}\;k'_{\sigma}\gamma^{\sigma}(1+\gamma_{5})v^{r}_{{\bf p},\mu}
\\\non &\times &
\;_{\nu}\langle 0(x^{0}_{I})
|\al^{t}_{{\bf k}-{\bf p},\nu_{\mu}}(z^{0})[\al^{v \dag}_{{\bf
q},\nu_{\mu}}(x^{0}) \al^{v }_{{\bf
q},\nu_{\mu}}(x^{0})-\beta^{v \dag}_{{\bf q},\nu_{\mu}}(x^{0})
\beta^{v }_{{\bf q},\nu_{\mu}}(x^{0})]\al^{u \dag}_{{\bf k'}-{\bf p},\nu_{\mu}}(y^{0}) |0(x^{0}_{I})\rangle_{\nu}
\\ &+& similar \; terms, \eea
(see Eq.(\ref{Echarge2}) in Appendix B).

\vspace{5mm}
\noindent \emph{Evaluation of terms like}
\vspace{5mm}

$_{\nu}\langle 0(x^{0}_{I})
|\al^{t}_{{\bf k}-{\bf p},\nu_{\mu}}(z^{0})\Big[\al^{v \dag}_{{\bf
q},\nu_{\mu}}(x^{0}) \al^{v }_{{\bf q},\nu_{\mu}}(x^{0})-
\beta^{v \dag}_{{\bf q},\nu_{\mu}}(x^{0}) \beta^{v }_{{\bf
q},\nu_{\mu}}(x^{0})\Big] \al^{u \dag}_{{\bf k'}-{\bf p},\nu_{\mu}}(y^{0})
|0(x^{0}_{I})\rangle_{\nu}$

\vspace{5mm}

We have
\bea \non &_{\nu}&\langle 0(x^{0}_{I}) |\al^{t}_{{\bf k}-{\bf p},
\nu_{\mu}}(z^{0})[\al^{v \dag}_{{\bf q}, \nu_{\mu}}(x^{0})
\al^{v }_{{\bf q},\nu_{\mu}}(x^{0})- \beta^{v \dag}_{{\bf
q},\nu_{\mu}}(x^{0}) \beta^{v }_{{\bf q},\nu_{\mu}}(x^{0})]
\al^{u \dag}_{{\bf k'}-{\bf p},\nu_{\mu}}(y^{0})
|0(x^{0}_{I})\rangle_{\nu}=
\\\non
\\\non &=&_{1,2}\langle 0
|G(x_{I}^{0})\al^{t}_{{\bf k}-{\bf p},\nu_{\mu}}(z^{0})
G^{-1}(x_{I}^{0})G(x_{I}^{0}) [\al^{v \dag}_{{\bf
q},\nu_{\mu}}(x^{0}) \al^{v }_{{\bf q},\nu_{\mu}}(x^{0})
\\\non &-&
\beta^{v \dag}_{{\bf q},\nu_{\mu}}(x^{0}) \beta^{v }_{{\bf
q},\nu_{\mu}}(x^{0})]G^{-1}(x_{I}^{0}) G(x_{I}^{0})\al^{u
\dag}_{{\bf k'}-{\bf p},\nu_{\mu}}(y^{0})G^{-1}(x_{I}^{0}) |0\rangle_{1,2}
\\\non
\\ \label{15} &=&_{1,2}\langle 0|\hat{\alpha}^{r}_{{\bf k}-{\bf p},\nu_{\mu}
}(x_{I}^{0},z^{0})[\hat{\al}^{v \dag}_{{\bf q},
\nu_{\mu}}(x_{I}^{0},x^{0}) \hat{\al}^{v }_{{\bf
q},\nu_{\mu}}(x_{I}^{0},x^{0})
- \hat{\beta}^{v \dag}_{{\bf
q},\nu_{\mu}}(x_{I}^{0},x^{0}) \hat{\beta}^{v }_{{\bf
q},\nu_{\mu}}(x_{I}^{0},x^{0})] \hat{\al}^{u \dag}_{{\bf k'}-{\bf p},
\nu_{\mu}}(x_{I}^{0},y^{0}) |0\rangle_{1,2}, \eea
Where the flavor annihilation operators are $\alpha^{r}_{{\bf
q},\nu_{\sigma}}(t,\theta)\equiv G^{-1}_{\bf \te}(t)\;\alpha
_{{\bf q},i}^{r}\;G_{\bf \te}(t)$:
\bea\label{ann3} \non \alpha^{r}_{{\bf
q},\nu_{e}}(t,\theta)&=&\cos\theta\;\alpha^{r}_{{\bf
q},1}\;+\;\sin\theta\;\sum_{s}\left[u^{r\dag}_{{\bf q},1}(t)
u^{s}_{{\bf q},2}(t)\; \alpha^{s}_{{\bf q},2}\;+\; u^{r\dag}_{{\bf
q},1}(t) v^{s}_{-{\bf q},2}(t)\; \beta^{s\dag}_{-{\bf q},2}\right]
\\ \non
\alpha^{r}_{{\bf
q},\nu_{\mu}}(t,\theta)&=&\cos\theta\;\alpha^{r}_{{\bf q},2}\;-
\;\sin\theta\;\sum_{s}\left[u^{r\dag}_{{\bf q},2}(t) u^{s}_{{\bf
q},1}(t)\; \alpha^{s}_{{\bf q},1}\;+\; u^{r\dag}_{{\bf q},2}(t)
v^{s}_{-{\bf q},1}(t)\; \beta^{s\dag}_{-{\bf q},1}\right]
\\ \non
\beta^{r}_{-{\bf
q},\nu_{e}}(t,\theta)&=&\cos\theta\;\beta^{r}_{-{\bf q},1}\;+
\;\sin\theta\;\sum_{s}\left[v^{s\dag}_{-{\bf q},2}(t) v^{r}_{-{\bf
q},1}(t)\; \beta^{s}_{-{\bf q},2}\;+\; u^{s\dag}_{{\bf q},2}(t)
v^{r}_{-{\bf q},1}(t)\; \alpha^{s\dag}_{{\bf q},2}\right]
\\ \non
\beta^{r}_{-{\bf
q},\nu_{\mu}}(t,\theta)&=&\cos\theta\;\beta^{r}_{-{\bf q},2}\;-
\;\sin\theta\;\sum_{s}\left[v^{s\dag}_{-{\bf q},1}(t) v^{r}_{-{\bf
q},2}(t)\; \beta^{s}_{-{\bf q},1}\;+\; u^{s\dag}_{{\bf q},1}(t)
v^{r}_{-{\bf q},2}(t)\; \alpha^{s\dag}_{{\bf q},1}\right].
\\\label{annih1}
\eea

By using
$\hat{\alpha}^{r}_{{\bf q},\nu_{\sigma} }(t',t) \equiv
G_{\bf \te}(t')\; \alpha^{r}_{{\bf q},\nu_{\sigma}}(t,\theta)\;
G^{-1}_{\bf \te}(t')$, and $G_{\bf \te}^{-1}(t) = G_{\bf - \te}(t)$,
we have
\bea\label{annih3} \non \hat{\alpha}^{r}_{{\bf q},\nu_{e} }(t',t)
&=&\cos\theta\; \alpha^{r}_{{\bf q},\nu_{e}}(t',-\theta)\;\\
\non &+& \;\sin\theta\;\sum_{s}\left[u^{r\dag}_{{\bf q},1}(t)
u^{s}_{{\bf q},2}(t)\; \alpha^{s}_{{\bf
q},\nu_{\mu}}(t',-\theta)\;
 + \; u^{r\dag}_{{\bf q},1}(t) v^{s}_{-{\bf q},2}(t)\;
\beta^{s\dag}_{-{\bf q,\nu_{\mu}}} (t',-\theta)\right]
\\ \non
\hat{\alpha}^{r}_{{\bf
q},\nu_{\mu}}(t',t)&=&\cos\theta\;\alpha^{r}_{{\bf
q},\nu_{\mu}}(t',-\theta)\;\\
\non &-& \;\sin\theta\;\sum_{s}
\left[u^{r\dag}_{{\bf q},2}(t) u^{s}_{{\bf q},1}(t)\;
\alpha^{s}_{{\bf q},\nu_{e}}(t',-\theta)\; +\; u^{r\dag}_{{\bf
q},2}(t) v^{s}_{-{\bf q},1}(t)\; \beta^{s\dag}_{-{\bf
q},\nu_{e}}(t',-\theta)\right]
\\ \non
\hat{\beta}^{r}_{-{\bf q},\nu_{e}}(t',t)&=&\cos\theta\;
\beta^{r}_{-{\bf q},\nu_{e}} (t',-\theta)\;\\
\non &+& \;\sin\theta\;\sum_{s}\left[v^{s\dag}_{-{\bf q},2}(t)
v^{r}_{-{\bf q},1}(t)\; \beta^{s}_{-{\bf
q},\nu_{\mu}}(t',-\theta)\;+\; u^{s\dag}_{{\bf q},2}(t)
v^{r}_{-{\bf q},1}(t)\; \alpha^{s\dag}_{{\bf
q},\nu_{\mu}}(t',-\theta)\right]
\\ \non
\hat{\beta}^{r}_{-{\bf
q},\nu_{\mu}}(t',t)&=&\cos\theta\;\beta^{r}_{-{\bf
q},\nu_{\mu}}(t',-\theta)\;
\\
 &-& \;\sin\theta\;\sum_{s}\left[v^{s\dag}_{-{\bf q},1} (t)
v^{r}_{-{\bf q},2}(t)\; \beta^{s}_{-{\bf
q},\nu_{e}}(t',-\theta)\;+\; u^{s\dag}_{{\bf q},1}(t) v^{r}_{-{\bf
q},2}(t)\; \alpha^{s\dag}_{{\bf q},\nu_{e}}(t',-\theta)\right].
\eea
By using Eqs.(\ref{annih1}) in Eqs.(\ref{annih3}) the last
equality in Eqs.(\ref{15})  can be finally expressed in terms of
the massive neutrino operators $\alpha^{r}_{{\bf q},i}$,
$\alpha^{r, \dag}_{{\bf q},i}$, $\beta^{r}_{-{\bf q},i}$,
$\beta^{r, \dag}_{-{\bf q},i}$ which act on the massive
neutrino vacuum $|0\rangle_{1,2}$.
Thus our computation allows the expression of the matrix element
 $\langle \Pi(t) | Q_{\nu_{\sigma}} | \Pi(t) \rangle$
as a function of states that have the proper mathematical characterization of the LSZ formalism.

We observe that the formula (\ref{CHARGE1}) is consistent with Eq.(14) of
Ref.\cite{Fujii:2004yd}: indeed, when the pion state is represented by a plane wave
(as done in Ref.\cite{Fujii:2004yd}),
the two equations acquire the same form, except for the neutrino charge expectation value,
which in our case includes the flavor vacuum contributions.
If we neglect such flavor vacuum effect,
we precisely recover Eq.(14) of Ref.\cite{Fujii:2004yd}.

Furthermore, when we consider the effect of finite lifetime of the pion and we take the neutrino
oscillation time $x^0 - x^0_I$ much bigger than $1/\Ga_\pi$, then $\al^{r \dag}_{{\bf k},\nu_{\mu}}(y^{0})
|0(x^{0}_{I})\rangle_{\nu} \simeq |\nu^{r}_{\;{\bf k},\mu}(x^{0}_{I})\rangle$  and
we obtain a form which is similar (once integrated in time) to
the one considered in Ref.\cite{spaceosc} giving the space dependent oscillation formula, except for the fact that
the pion state rather than the neutrino state is represented by a wave packet.
The final expression ensuing the formula (\ref{CHARGE1}) is too lengthy to be discussed at this stage.
Further analysis will be done elsewhere.

\section{Conclusion}

In this paper we have studied neutrino mixing and oscillations in quantum field
theory and we discussed the determination of the oscillation probability
including the neutrino production vertex.
A crucial point in our analysis is the disclosure of the fact that in order
to describe the neutrino oscillations we have to use
the flavor states defined as
$|\nu^{r}_{\;{\bf k},\si}\ran = \al^{r\dag}_{{\bf k},{\nu_{\sigma}}}(0) |0(0)\ran_{\nu}$,
with $\si = e,\mu,\tau $ and the flavor charge operators $\nof\wwQ_{\nu_{\sigma}}\nof$ normal
ordered with respect to the flavor vacuum.
Indeed, we have shown that the usual Pontecorvo states
are not eigenstates of the flavor charges $: \wQ_{\nu_{\sigma}}:$ and $\nof\wwQ_{\nu_{\sigma}}\nof$.

We showed that the possibility of different mass parameters is
intrinsic to the very same structure of QFT and is independent of
the occurrence or not of the field mixing; hence, as noted in
\cite{Blasone:1999jb,JM011,yBCV02},
the mass parameters must be chosen not arbitrarily,
but on the ground of physical reasons: $\mu_e \equiv m_1$, $\mu_\mu \equiv m_2$,
$\mu_\tau \equiv m_3$.

Moreover, we have computed the neutrino lepton charge in decay processes
where neutrinos are generated, proving
that the corresponding lepton charge expectation value
can be uniquely expressed in terms of LSZ states and in particular of
the massive neutrino annihilation/creation operators acting on the
massive neutrino vacuum.

\section*{Acknowledgements}

The present research has been carried out in the framework of the
ESF network COSLAB and we acknowledge partial financial support
by INFN, INFM and MIUR.

\appendix

\section{Orthogonality of flavor states at different times}

The product of two vacuum states at different times $t\neq t'$ (we
put for simplicity $t'=0$) is
\begin{eqnarray}
{}_{\nu}\langle0|0(t)\rangle_{\nu} = \prod_{k} C_{k}^{2}(t)=
e^{2\sum_{k}ln C_{k}(t)}
\end{eqnarray}
with
\begin{eqnarray}\nonumber C_{k}(t) &\equiv&
(1-\sin^{2}\theta\;|V_{k}|^{2})^{2}
+\;2\;\sin^{2}\theta\;\cos^{2}\theta\; |V_{k}|^{2} \;  e^{-i
(\omega_{k,2} + \omega_{k,1}) t}+
 \\
\label{bit} &+&\;\sin^{4}\theta \;|V_{k}|^{2} \;|U_{k}|^{2}
\left(e^{- 2 i \omega_{k,1} t} \; + \;e^{-2 i \omega_{k,2} t}\;
\right) +\sin^{4}\theta \; |V_{k}|^{4} \; e^{- 2 i (\omega_{k,2} +
\omega_{k,1}) t}\;.
\end{eqnarray}
In the infinite volume limit we obtain (note that $|C_{k}(t)| \leq
1$ for any value of $k$, $t$, and of the parameters $\theta$, $
m_1$, $m_2$ ):
\begin{equation}\label{lim}
\lim_{V \rightarrow
\infty}\;{}_{\nu}\langle0|0(t)\rangle_{\nu} = \lim_{V
\rightarrow \infty}\; exp\left[\frac{2V}{(2\pi)^{3}}\int d^{3}{\bf k} \;
\left( ln\;|C_{k}(t)|\; + \; i \alpha_{k}(t) \right)\right]=0
\end{equation}
 with $|C_{k}(t)|^{2}= Re[C_{k}(t)]^{2} + Im[C_{k}(t)]^{2} $ and
 $\alpha_{k}(t)= \tan^{-1}\left(Im[C_{k}(t)]/ Re[C_{k}(t)]
 \right)$.

Thus we have orthogonality of the vacua at different times\footnote{Note that it may
be that for some values of the continuous index $k$, the $C_k(t)$ are periodic functions
of time. However, the set of values of $k$ for which this happens is a zero-measure set in the
above integration Eq.(\ref{lim}).}.

One can easily check that flavor neutrino states are also orthogonal at
different times.
Consider the electron neutrino state at time $t$ with momentum
${\bf k}$:
\bea |\nu^{r}_{{\bf k},e}(t)\rangle = \alpha^{r \dag}_{{\bf k},\nu_{e}}(t)
|0(t)\rangle_{\nu}.
\eea
The flavor vacuum is explicitly given by
\bea |0(t)\rangle_{\nu}= \prod_{\bf p}G^{-1}_{{\bf
p},\theta}(t)|0\rangle_{1,2}, \eea
where, to be precise, the mass vacuum is to be understood as
$|0\rangle_{1,2}= |0\rangle^{{\bf k}_{1}}_{1,2}\bigotimes
|0\rangle^{{\bf k}_{2}}_{1,2}\bigotimes |0\rangle^{{\bf
k}_{3}}_{1,2}... $.

Then, we have
\bea \langle \nu^{r}_{{\bf k},e}(0)|\nu^{r}_{{\bf k},e}(t)\rangle &=&
{}_{\nu}\langle0|\alpha^{r }_{{\bf k},\nu_{e}}(0)\alpha^{r \dag}_{{\bf
k},\nu_{e}}(t)|0(t)\rangle_{\nu}\\ &=& \left(\prod_{\bf p}
{}_{1,2}\langle 0|G_{{\bf p},\theta}(0)\right)\alpha^{r }_{{\bf
k},\nu_{e}}(0)\alpha^{r \dag}_{{\bf k},\nu_{e}}(t)\left(\prod_{\bf
q}\;G^{-1}_{{\bf q},\theta}(t)|0\rangle_{1,2}\right), \eea
and, since for ${\bf p}\neq{\bf q}$ the mixing generators commute
among themselves and with $\alpha^{r }_{{\bf k},\nu_{e}}$ for ${\bf
k}\neq {\bf p}, {\bf q}$, it is
\bea \langle \nu^{r}_{{\bf k},e}(0)|\nu^{r}_{{\bf k},e}(t)\rangle &\propto
&\;{}_{\nu}\langle 0^{\bf k}|\alpha^{r }_{{\bf k},\nu_{e}}(0)\alpha^{r
\dag}_{{\bf k},\nu_{e}}(t)|0^{\bf k}(t)\rangle_{\nu} \prod_{\bf p\neq
k}\; {}_{1,2}\langle 0 |G_{{\bf p},\theta}(0)G^{-1}_{{\bf
p},\theta}(t)|0\rangle_{1,2}
\\&=&
 {}_{\nu}\langle 0^{\bf k}|\alpha^{r }_{{\bf
k},\nu_{e}}(0)\alpha^{r \dag}_{{\bf k},\nu_{e}}(t)|0^{\bf
k}(t)\rangle_{\nu}\;{}_{\nu}\langle0|0(t)\rangle_{\nu}.
\eea
By using Eq.(\ref{lim}),  we obtain the orthogonality of flavor
states at different times in the infinite volume limit, provided
${}_{\nu}\langle 0^{\bf k}|\alpha^{r }_{{\bf k},\nu_{e}}(0)\alpha^{r
\dag}_{{\bf k},\nu_{e}}(t)|0^{\bf k}(t)\rangle_{\nu}$ is finite or
zero, as it is indeed.

\section{Explicit expressions of Eqs.(\ref{charge}), (\ref{charge1}), (\ref{charge2})}

The complete Eq.(\ref{CHARGE}) is:

\bea \label{Echarge}\non
\non \langle \nof\wwQ_{\nu_{\mu}}(x^{0})\nof\rangle &=&
-|g_{\pi}|^{2} \int^{x^{0}}_{x^{0}_{I}}
\int^{x^{0}}_{x^{0}_{I}}d^{4}z d^{4}y \langle \pi(\langle{\bf
k}\rangle, {\bf x}, x^{0}_{I})|
\partial_{\lambda}\pi^{\dag}(z)
\partial_{\sigma}\pi(y)| \pi(\langle{\bf k}\rangle, {\bf x}, x^{0}_{I})
\rangle\;
\\\non&\times &\sum_{r,t,u,v}\int \frac{d^{3}{\bf p}\;
d^{3}{\bf q}\; d^{3}{\bf q'}\; d^{3}{\bf q''}}{(2\pi)^{6}}\;
\Big(e^{i({\bf p}+{\bf q})\cdot{\bf z}} e^{-i({\bf p}+{\bf
q'})\cdot {\bf y}}
\\\non &\times &
e^{-i(\omega_{p}+\omega_{q,2})z^{0}}
e^{i(\omega_{p}+\omega_{q',2})y^{0}} \bar{v}^{r}_{{\bf
p},\mu}\gamma^{\lambda}(1+\gamma_{5}) u^{t}_{{\bf
q},2}\bar{u}^{u}_{{\bf q'},2}\gamma^{\sigma}(1+\gamma_{5})
v^{r}_{{\bf p},\mu}
\\\non &\times & \;_{\nu}\langle 0(x^{0}_{I})
|\al^{t}_{{\bf q},\nu_{\mu}}(z^{0})[\al^{v \dag}_{{\bf
q''},\nu_{\mu}}(x^{0}) \al^{v }_{{\bf
q''},\nu_{\mu}}(x^{0})-\beta^{v \dag}_{{\bf q''},\nu_{\mu}}(x^{0})
\beta^{v }_{{\bf q''},\nu_{\mu}}(x^{0})]\al^{u \dag}_{{\bf
q'},\nu_{\mu}}(y^{0}) |0(x^{0}_{I})\rangle_{\nu}
\\\non &+& e^{i({\bf p}-{\bf q})\cdot{\bf z}}
e^{-i({\bf p}-{\bf q'})\cdot {\bf y}}
e^{-i(\omega_{p}-\omega_{q,2})z^{0}}
e^{i(\omega_{p}-\omega_{q',2})y^{0}} \bar{v}^{r}_{{\bf
p},\mu}\gamma^{\lambda}(1+\gamma_{5}) v^{t}_{{\bf
q},2}\bar{v}^{u}_{{\bf q'},2}\gamma^{\sigma}(1+\gamma_{5})
v^{r}_{{\bf p},\mu}
\\\non &\times & \;_{\nu}\langle 0(x^{0}_{I})
|\beta^{t \dag}_{{\bf q},\nu_{\mu}}(z^{0})[\al^{v \dag}_{{\bf
q''},\nu_{\mu}} (x^{0}) \al^{v }_{{\bf
q''},\nu_{\mu}}(x^{0})-\beta^{v \dag}_{{\bf q''},\nu_{\mu}}
(x^{0}) \beta^{v }_{{\bf q''},\nu_{\mu}}(x^{0})]\beta^{u}_{{\bf
q'},\nu_{\mu}}(y^{0}) |0(x^{0}_{I})\rangle_{\nu}
\\\non &+& e^{i({\bf p}+{\bf q})\cdot{\bf z}}
e^{-i({\bf p}-{\bf q'})\cdot {\bf y}}
e^{-i(\omega_{p}+\omega_{q,2})z^{0}}
e^{i(\omega_{p}-\omega_{q',2})y^{0}} \bar{v}^{r}_{{\bf
p},\mu}\gamma^{\lambda}(1+\gamma_{5}) u^{t}_{{\bf
q},2}\bar{v}^{u}_{{\bf q'},2}\gamma^{\sigma}(1+\gamma_{5})
v^{r}_{{\bf p},\mu}
\\\non &\times & \;_{\nu}\langle 0(x^{0}_{I})
|\al^{t}_{{\bf q},\nu_{\mu}}(z^{0})[\al^{v \dag}_{{\bf
q''},\nu_{\mu}}(x^{0}) \al^{v }_{{\bf
q''},\nu_{\mu}}(x^{0})-\beta^{v \dag}_{{\bf q''},\nu_{\mu}}
(x^{0}) \beta^{v }_{{\bf q''},\nu_{\mu}}(x^{0})]\beta^{u}_{{\bf
q'},\nu_{\mu}}(y^{0}) |0(x^{0}_{I})\rangle_{\nu}
\\\non &+& e^{i({\bf p}-{\bf q})\cdot{\bf z}}
e^{-i({\bf p}+{\bf q'})\cdot {\bf y}}
e^{-i(\omega_{p}-\omega_{q,2})z^{0}}
e^{i(\omega_{p}+\omega_{q',2})y^{0}} \bar{v}^{r}_{{\bf
p},\mu}\gamma^{\lambda}(1+\gamma_{5}) v^{t}_{{\bf
q},2}\bar{u}^{u}_{{\bf q'},2}\gamma^{\sigma}(1+\gamma_{5})
v^{r}_{{\bf p},\mu}
\\ &\times & \;_{\nu}\langle 0(x^{0}_{I})
|\beta^{t \dag}_{{\bf q},\nu_{\mu}}(z^{0})[\al^{v \dag}_{{\bf
q''},\nu_{\mu}} (x^{0}) \al^{v }_{{\bf
q''},\nu_{\mu}}(x^{0})-\beta^{v \dag}_{{\bf q''},\nu_{\mu}}
(x^{0}) \beta^{v }_{{\bf q''},\nu_{\mu}}(x^{0})] \al^{u
\dag}_{{\bf q'},\nu_{\mu}}(y^{0}) |0(x^{0}_{I})\rangle_{\nu}\Big)
 \eea

The complete Eq.(\ref{CHARGE1}) is:
\bea \non\label{Echarge1} \non \langle \nof\wwQ_{\nu_{\mu}}(x^{0})\nof\rangle &=&
-|g_{\pi}|^{2} \int^{x^{0}}_{x^{0}_{I}}
\int^{x^{0}}_{x^{0}_{I}}d^{4}z d^{4}y \sum_{r,t,u,v}\int
\frac{d^3{\bf k}\; d^3{\bf k'} d^{3}{\bf p}\; d^{3}{\bf q}\;
d^{3}{\bf q'}\; d^{3}{\bf q''}}{4(2\pi)^{9}{\Omega_{ k}\Omega_{
k'}}}\; A^{\ast}_{\pi}({\bf k},\langle{\bf k}\rangle)
A_{\pi}({\bf k'},\langle{\bf k'}\rangle)
\\\non&\times & k_{\lambda}k'_{\sigma}
e^{i ({\bf k}-{\bf k'}) \cdot {\bf x}}  e^{-i
(\Omega_{k}-\Omega_{k'})x^{0}_{I}}
\\\non&\times &
\Big(e^{-i({\bf k}-{\bf p}-{\bf q})\cdot{\bf z}} e^{i({\bf
k'}-{\bf p}-{\bf q'})\cdot {\bf y}}
e^{i(\Omega_{ k}-\omega_{p}-\omega_{q,2})(z^{0}-x^{0}_{I})}
e^{-i(\Omega_{ k'}-\omega_{p}-\omega_{q',2})(y^{0}-x^{0}_{I})}
\bar{v}^{r}_{{\bf
p},\mu}\gamma^{\lambda}(1+\gamma_{5}) u^{t}_{{\bf
q},2}\bar{u}^{u}_{{\bf q'},2}
\\\non &\times & \gamma^{\sigma}(1+\gamma_{5})v^{r}_{{\bf p},\mu}
\;_{\nu}\langle 0(x^{0}_{I})
|\al^{t}_{{\bf q},\nu_{\mu}}(z^{0})[\al^{v \dag}_{{\bf
q''},\nu_{\mu}}(x^{0}) \al^{v }_{{\bf
q''},\nu_{\mu}}(x^{0})-\beta^{v \dag}_{{\bf q''},\nu_{\mu}}(x^{0})
\beta^{v }_{{\bf q''},\nu_{\mu}}(x^{0})]\al^{u \dag}_{{\bf
q'},\nu_{\mu}}(y^{0}) |0(x^{0}_{I})\rangle_{\nu}
\\\non &+& e^{-i({\bf k}-{\bf p}+{\bf q})\cdot{\bf z}}
e^{i({\bf k'}-{\bf p}+{\bf q'})\cdot {\bf y}}
e^{i(\Omega_{ k}-\omega_{p}+\omega_{q,2})(z^{0}-x^{0}_{I})}
e^{-i(\Omega_{ k'}-\omega_{p}+\omega_{q',2})(y^{0}-x^{0}_{I})}
 \bar{v}^{r}_{{\bf
p},\mu}\gamma^{\lambda}(1+\gamma_{5}) v^{t}_{{\bf
q},2}\bar{v}^{u}_{{\bf q'},2}
\\\non &\times & \gamma^{\sigma}(1+\gamma_{5})
v^{r}_{{\bf p},\mu}\;_{\nu}\langle 0(x^{0}_{I})
|\beta^{t \dag}_{{\bf q},\nu_{\mu}}(z^{0})[\al^{v \dag}_{{\bf
q''},\nu_{\mu}} (x^{0}) \al^{v }_{{\bf
q''},\nu_{\mu}}(x^{0})-\beta^{v \dag}_{{\bf q''},\nu_{\mu}}
(x^{0}) \beta^{v }_{{\bf q''},\nu_{\mu}}(x^{0})]\beta^{u}_{{\bf
q'},\nu_{\mu}}(y^{0}) |0(x^{0}_{I})\rangle_{\nu}
\\\non &+& e^{-i({\bf k}-{\bf p}-{\bf q})\cdot{\bf z}}
e^{i({\bf k'}-{\bf p}+{\bf q'})\cdot {\bf y}}
e^{i(\Omega_{ k}-\omega_{p}-\omega_{q,2})(z^{0}-x^{0}_{I})}
e^{-i(\Omega_{ k'}-\omega_{p}+\omega_{q',2})(y^{0}-x^{0}_{I})}
\bar{v}^{r}_{{\bf
p},\mu}\gamma^{\lambda}(1+\gamma_{5}) u^{t}_{{\bf
q},2}\bar{v}^{u}_{{\bf q'},2}
\\\non &\times & \gamma^{\sigma}(1+\gamma_{5})
v^{r}_{{\bf p},\mu}\;_{\nu}\langle 0(x^{0}_{I})
|\al^{t}_{{\bf q},\nu_{\mu}}(z^{0})[\al^{v \dag}_{{\bf
q''},\nu_{\mu}}(x^{0}) \al^{v }_{{\bf
q''},\nu_{\mu}}(x^{0})-\beta^{v \dag}_{{\bf q''},\nu_{\mu}}
(x^{0}) \beta^{v }_{{\bf q''},\nu_{\mu}}(x^{0})]\beta^{u}_{{\bf
q'},\nu_{\mu}}(y^{0}) |0(x^{0}_{I})\rangle_{\nu}
\\\non &+& e^{-i({\bf k}-{\bf p}+{\bf q})\cdot{\bf z}}
e^{i({\bf k'}-{\bf p}-{\bf q'})\cdot {\bf y}}
e^{i(\Omega_{ k}-\omega_{p}+\omega_{q,2})(z^{0}-x^{0}_{I})}
e^{-i(\Omega_{ k'}-\omega_{p}-\omega_{q',2})(y^{0}-x^{0}_{I})}
 \bar{v}^{r}_{{\bf
p},\mu}\gamma^{\lambda}(1+\gamma_{5}) v^{t}_{{\bf
q},2}\bar{u}^{u}_{{\bf q'},2}
\\\non &\times & \gamma^{\sigma}(1+\gamma_{5})
v^{r}_{{\bf p},\mu}\;_{\nu}\langle 0(x^{0}_{I})
|\beta^{t \dag}_{{\bf q},\nu_{\mu}}(z^{0})[\al^{v \dag}_{{\bf
q''},\nu_{\mu}} (x^{0}) \al^{v }_{{\bf
q''},\nu_{\mu}}(x^{0})-\beta^{v \dag}_{{\bf q''},\nu_{\mu}}
(x^{0}) \beta^{v }_{{\bf q''},\nu_{\mu}}(x^{0})] \al^{u
\dag}_{{\bf q'},\nu_{\mu}}(y^{0}) |0(x^{0}_{I})\rangle_{\nu}\Big).
\\
 \eea

The complete Eq.(\ref{CHARGE2}) is:
\bea \non\label{Echarge2} \non \langle \nof\wwQ_{\nu_{\mu}}(x^{0})\nof\rangle &=&
-|g_{\pi}|^{2} \int^{x^{0}}_{x^{0}_{I}}
\int^{x^{0}}_{x^{0}_{I}}dz^{0} dy^{0} \sum_{r,t,u,v}\int
\frac{ d^{3}{\bf k}\;
d^{3}{\bf k'}\; d^{3}{\bf p}\;  d^{3}{\bf q}}{4(2\pi)^{3}{\Omega_{k}\Omega_{
k'}}}\; A^{\ast}_{\pi}({\bf k},\langle{\bf k}\rangle)
A_{\pi}({\bf k'},\langle{\bf k}\rangle)\;
e^{i ({\bf k}-{\bf k'}) \cdot {\bf x}}  e^{-i
(\Omega_{k}-\Omega_{k'})x^{0}_{I}}
\\\non&\times &
\Big(e^{i(\Omega_{k}-\omega_{p}-\omega_{k-p,2})(z^{0}-x^{0}_{I})}
e^{-i(\Omega_{k'}-\omega_{p}-\omega_{k'-p,2})(y^{0}-x^{0}_{I})}
\bar{v}^{r}_{{\bf
p},\mu}k_{\lambda}\gamma^{\lambda}(1+\gamma_{5}) u^{t}_{{\bf k}-{\bf p},2}
\bar{u}^{u}_{{\bf k'}-{\bf p},2}\;k'_{\sigma}\gamma^{\sigma}(1+\gamma_{5})v^{r}_{{\bf p},\mu}
\\\non &\times &
\;_{\nu}\langle 0(x^{0}_{I})
|\al^{t}_{{\bf k}-{\bf p},\nu_{\mu}}(z^{0})[\al^{v \dag}_{{\bf
q},\nu_{\mu}}(x^{0}) \al^{v }_{{\bf
q},\nu_{\mu}}(x^{0})-\beta^{v \dag}_{{\bf q},\nu_{\mu}}(x^{0})
\beta^{v }_{{\bf q},\nu_{\mu}}(x^{0})]\al^{u \dag}_{{\bf k'}-{\bf p},\nu_{\mu}}(y^{0}) |0(x^{0}_{I})\rangle_{\nu}
\\\non &+&
e^{i(\Omega_{k }-\omega_{p}+\omega_{p-k,2})(z^{0}-x^{0}_{I})}
e^{-i(\Omega_{k'}-\omega_{p}+\omega_{p-k',2})(y^{0}-x^{0}_{I})}
 \bar{v}^{r}_{{\bf
p},\mu}k_{\lambda}\gamma^{\lambda}(1+\gamma_{5}) v^{t}_{{\bf
p-k},2}\bar{v}^{u}_{{\bf p-k'},2}\; k'_{\sigma}\gamma^{\sigma}(1+\gamma_{5})
v^{r}_{{\bf p},\mu}
\\\non &\times & \;_{\nu}\langle 0(x^{0}_{I})
|\beta^{t \dag}_{{\bf p}-{\bf k},\nu_{\mu}}(z^{0})[\al^{v \dag}_{{\bf
q},\nu_{\mu}} (x^{0}) \al^{v }_{{\bf
q},\nu_{\mu}}(x^{0})-\beta^{v \dag}_{{\bf q},\nu_{\mu}}
(x^{0}) \beta^{v }_{{\bf q},\nu_{\mu}}(x^{0})]\beta^{u}_{{\bf p}-{\bf k'},
\nu_{\mu}}(y^{0}) |0(x^{0}_{I})\rangle_{\nu}
\\\non &+&
e^{i(\Omega_{k}-\omega_{p}-\omega_{k-p,2})(z^{0}-x^{0}_{I})}
e^{-i(\Omega_{k'}-\omega_{p}+\omega_{p-k',2})(y^{0}-x^{0}_{I})}
\bar{v}^{r}_{{\bf
p},\mu}k_{\lambda}\gamma^{\lambda}(1+\gamma_{5}) u^{t}_{{\bf
k}-{\bf p},2}\bar{v}^{u}_{{\bf p}-{\bf k'},2}\; k'_{\sigma}\gamma^{\sigma}(1+\gamma_{5})
v^{r}_{{\bf p},\mu}
\\\non &\times &\;_{\nu}\langle 0(x^{0}_{I})
|\al^{t}_{{\bf k}-{\bf p},\nu_{\mu}}(z^{0})[\al^{v \dag}_{{\bf
q},\nu_{\mu}}(x^{0}) \al^{v }_{{\bf
q},\nu_{\mu}}(x^{0})-\beta^{v \dag}_{{\bf q},\nu_{\mu}}
(x^{0}) \beta^{v }_{{\bf q},\nu_{\mu}}(x^{0})]\beta^{u}_{{\bf p}-{\bf
k'},\nu_{\mu}}(y^{0}) |0(x^{0}_{I})\rangle_{\nu}
\\\non &+&
e^{i(\Omega_{k}-\omega_{p}+\omega_{p-k,2})(z^{0}-x^{0}_{I})}
e^{-i(\Omega_{k'}-\omega_{p}-\omega_{k'+p,2})(y^{0}-x^{0}_{I})}
 \bar{v}^{r}_{{\bf
p},\mu}k_{\lambda}\gamma^{\lambda}(1+\gamma_{5}) v^{t}_{{\bf
p}-{\bf
k},2}\bar{u}^{u}_{{\bf k'}+{\bf
p},2}\; k'_{\sigma}\gamma^{\sigma}(1+\gamma_{5})
v^{r}_{{\bf p},\mu}
\\ \non &\times & \;_{\nu}\langle 0(x^{0}_{I})
|\beta^{t \dag}_{{\bf
p}-{\bf
k},\nu_{\mu}}(z^{0})[\al^{v \dag}_{{\bf
q},\nu_{\mu}} (x^{0}) \al^{v }_{{\bf
q},\nu_{\mu}}(x^{0})-\beta^{v \dag}_{{\bf q},\nu_{\mu}}
(x^{0}) \beta^{v }_{{\bf q},\nu_{\mu}}(x^{0})] \al^{u
\dag}_{{\bf k'}+{\bf
p},\nu_{\mu}}(y^{0}) |0(x^{0}_{I})\rangle_{\nu}\Big).
\\ \eea

\end{document}